\documentclass[%
 reprint,
 amsmath,amssymb,
 aps,
 pra,
]{revtex4-1}

\usepackage{graphicx}
\usepackage{dcolumn}
\usepackage{bm}

\begin{document}

\title{Maximization of the thermoelectric cooling of graded Peltier by analytical heat equation resolution.}

\author{E. Thi\'ebaut$^1$}

\author{C. Goupil$^2$}

\author{F. Pesty$^1$}

\author{Y. D'Angelo$^3$}

\author{G. Guegan$^4$}

\author{P. Lecoeur$^1$}

\affiliation{$^1$Centre de Nanosciences et de Nanotechnologies, CNRS, Univ. Paris-Sud, Universit\'e Paris-Saclay, C2N-Orsay, 91405 Orsay, France \\
$^2$Laboratoire Interdisciplinaire des Energies de Demain (LIED), UMR 8236 Universit\'e Paris Diderot, CNRS, 5 Rue Thomas Mann, 75013 Paris, France \\
$^3$Laboratoire de Math\'ematiques J.A. Dieudonn\'e, Universit\'e de Nice Sophia Antipolis, CNRS UMR 7351,  Parc Valrose 06108 NICE CEDEX, France \\
$^4$STMicroelectronics TOURS, 10 Rue Thales de Milet, CS97155, 37071 Tours Cedex 2, France}

\begin{abstract}

Increasing the maximum cooling effect of a Peltier cooler can be achieved through materials and device design. The use of inhomogeneous, FGM (functionally graded materials) may be adopted in order to increase maximum cooling without improvement of the zT (figure of merit), however these systems are usually based on the assumption that the local optimization of the zT is the suitable criterion to increase thermoelectric performances. In the present paper, we solved the heat equation in a graded material and performed both analytic and numerical analysis of a graded Peltier cooler. We find a local criterion that we used to assess the possible improvement of graded materials for thermoelectric cooling. A fair improvement of cooling effect is predicted for semiconductor materials (up to $36\%$) and the best graded system for cooling is described. The influence of the equation of state of the electronic gas of the material is discussed, and the difference in term of entropy production between the graded and the classical system is also described.

\end{abstract}

\maketitle

\section*{Introduction}

Thermoelectric materials are used to promote energy harvesting and build cooling devices using direct thermoelectric conversion with no moving parts. Different parameters can be used to evaluate the performance of a thermoelectric system. For cooling applications, an important parameter is the maximum cooling temperature. A thermoelectric system made of homogeneous materials can impose a maximum cooling limited by its figure of merit ($zT$) and equal to $\Delta T = zT_{c}^{2}/2  $. The figure of merit is defined as $zT_{c} = \alpha^{2} T_{c} / (\kappa \rho) $, where $\alpha $ is the Seebeck coefficient, $ \kappa$ the thermal conductivity, $\rho $ the electrical resistivity and $ T_{c}$ the temperature of the cold side.

In thermoelectric systems based on the constant properties model (CPM) the Peltier effect is localized at the interface and the Joule effect is homogeneously spread across the device. Inhomogeneity leads to the Peltier-Thomson or extrinsic Peltier effect within the material and gives rise to inhomogeneous Joule heating. In a real device the temperature dependence of the thermoelectric material properties leads to such inhomogeneities. The CPM is therefore realistic only for small temperature differences. However the simplicity of such a system allows it to be easily used as comparison with more realistic systems. Resolution of inhomogeneous thermoelectric systems is complex and numerical computation of segmented systems with temperature dependence properties were performed in the 60's \cite{moore1962exact}. At the same time, material conditions for a segmented device to improve thermoelectric performance were investigated \cite{ure1962materials}. Beyond the segmented device, the graded device has been investigated with analytic resolution in the 60's for linear variation of the Seebeck coefficient and constant figure of merit \cite{ybarrondo1965influence}. Graded and segmented thermoelectric devices are sufficiently promising to be patented \cite{kountz1971thermoelectric}. With improved computer capabilities, numerical research on optimal solution for segmented \cite{buist1995extrinsic,schilz1998local,helmers1998graded} and graded \cite{mahan1991inhomogeneous} systems were performed during the 90's.

The performance of a thermoelectric system depends on the current density going through the material. In a segmented device the current that optimizes each segment might be different. The compatibility approach ("$u=s$") is used to optimize a thermoelectric system through the concept of reduced current ($u=$ the electrical flux divided by the heat flux)\cite{ursell2002compatibility}. The optimization is obtained when $u$ is equal to $s$ which depends on the material properties. The efficiency of a thermoelectric generator (TEG) and the coefficient of performance of a thermoelectric cooler (TEC) has been investigated using this approach for segmented devices \cite{ursell2002compatibility,snyder2003design,snyder2003thermoelectric,snyder2004application} and for graded devices \cite{seifert2008local,snyder2012improved,seifert2012exact,seifert2013self,seifert2014thermoelectric}. The reduced current is the local version of the Prandtl number \cite{apertet2012internal}.

On the experimental side, segmented thermoelectric generators based on $\mathrm{Bi_{2}Te_{3}}$ materials have been designed and measured \cite{vikhor2009generator,anatychuk2011segmented}. These works lead to a $15\%$ improvement of the generator efficiency.

However, unlike graded devices, segmented devices present the disavantage that their contact resistances can reduce the performance of the system \cite{vikhor2006theoretical}. The fabrication of graded material can be achieved for example by alloying silicon and germanium \cite{hedegaard2014functionally}.

The compatibility approach has been proven to be a powerful tool to optimize efficiency of TEG and coefficient of performance of TEC, however it has been shown that optimizing the maximum temperature difference requires another approach as discussed in \cite{seifert2014thermoelectric}. The temperature difference and the efficiency are different optimization targets.

The maximum temperature difference that a thermoelectric cooler can impose is a key parameter of the thermoelectric performance of a device. This parameter has been investigated through different approaches.

A hypothetical material where $zT$ stays constant as the Seebeck coefficient varies due to doping has been investigated by numerical \cite{muller2006separated} and analytical means \cite{bian2006beating}. Numerical calculations of the maximum temperature difference based on experimental material properties \cite{bian2006cooling, bian2007maximum} yield theoretical improvements of $27\%$ for $\mathrm{Bi_{2}Te_{3}}$ materials and $35\%$ for silicon.

In the present paper, we provide an analytic solution for a graded thermoelectric system, maximizing the temperature difference in the case of a general expression of the Seebeck coefficient as well as the electric conductivity as functions of the doping level. The analytic solution is studied by analytic and numerical means in the case of a simple semiconductor model.

In the first section we present an analytical analysis of a graded material in a thermoelectric cooler. The optimization of the temperature maximum is based on the work initiated by Bian et al. \cite{bian2006beating}. We show a local criterion that can be established to deduce the best graded material.

In the second section, the relation between the Seebeck coefficient and the electric conductivity is discussed as a consequence of the equation of state of the material in use. A graded system can be manufactured through different methods (doped semiconductors, alloy) and materials (silicon, oxides, classic thermoelectric materials, polymers). Each method or material leads to a different equation of state relating the Seebeck coefficient to the electrical conductivity.

As most good thermoelectrics are semiconductors, a good way to obtain a graded material should be to use a graded doped semiconductor. In the third section we apply the local criterion for a simple thermoelectric model of a semiconductor in order to evaluate the doping level as a function of the position and to calculate the improvement of the cooling effect.

From our optimization, the doping level maximizing the cooling is a linear profile. We analyzed the influence of a linear doping level in a semiconductor. Under these conditions the cooling as a function of the Seebeck coefficient at the cold side and the Seebeck coefficient at the hot side is plotted and discussed.

In the fourth section, to evaluate the validity of the analytical solution, the system is numerically computed and differences between numerical and analytical results are presented. However the results validate the trend arising from the analytical solution. An evaluation of the entropy sources is performed based on the numerical results.

\section{Analytical optimization of a graded thermoelectric material}

The heat equation in a one-dimensional graded thermoelectric system in static regime includes three terms, respectively : heat conduction, Thomson effect and Joule effect: 

\begin{eqnarray}
0 = \frac{\partial}{\partial x} (\kappa (x,T(x)) \frac{\partial T}{\partial x}(x)) \nonumber\\
- J T(x) \frac{\partial \alpha}{\partial x}(x,T(x)) + \rho (x,T(x)) J^{2} 
\label{eq chaleur fourrier}
\end{eqnarray}

Eq. (\ref{eq chaleur fourrier}) in the general situation cannot be solved analytically. If we assume that the thermal conductivity is constant, the material properties are independent from temperature variations, and the temperature stays close to $T_{0}$ ($J T(x) \frac{\partial \alpha}{\partial x}(x) = JT_{0}\frac{\partial \alpha}{\partial x}(x)$) we obtain:

\begin{eqnarray}
0 = \kappa \frac{\partial^{2} T }{\partial x^{2}} (x)  - J T_{0} \frac{\partial \alpha}{\partial x}(x) + \rho (x) J^{2} 
\label{eq chaleur fourrier simplifie}
\end{eqnarray}

The system that is considered is composed of one $n$ thermoelectric material from $x=-L$ to $x=0$ and one $p$ thermoelectric material from $x=0$ to $x=L$. At the position $x=-L$ and $x=L$ the system is in contact with a thermostat. This is a symmetric system for the thermal conductivity ($\kappa$) and the electrical resistivity ($\rho$), and antisymmetric for the Seebeck coefficient ($\alpha$). This system models a Peltier cooler where the cold side ($x=0$) is at the interface between the $n$ and the $p$ thermoelectric element.   

By integrating twice Eq. (\ref{eq chaleur fourrier simplifie}) the maximal temperature can be computed and the current density can be optimized as shown in \cite{bian2006beating} to obtain the maximum temperature difference :

\begin{eqnarray}
\Delta T = \frac{T_{0}^{2}}{4\kappa} \frac{ \left( \int_0^L \alpha(v) \, \mathrm dv  \right) ^{2}  }{ \int_0^L  \int_0^v \rho(u) \, \mathrm du \, \mathrm dv } = \frac{A}{B} \frac{T_{0}^{2}}{4\kappa}
\label{dt pc max}
\end{eqnarray}

Eq.(\ref{dt pc max}) was obtained and analysed in \cite{bian2006beating} and it has been shown that a graded system can improve the maximum cooling a Peltier cooler can reach.

\begin{eqnarray}
B =  \int_0^L  \int_0^v  \rho(u)  \, \mathrm du \, \mathrm dv  
\end{eqnarray}

\begin{eqnarray}
A =  \left( \int_0^L \alpha(v) \, \mathrm dv  \right) ^{2}    
\end{eqnarray}

The temperature difference depends on the function of the Seebeck coefficient ($\alpha (x)$) as a function of the position ($x$) and on the function of the electrical resistivity ($\rho (x)$) as a function of the position. Maximization of $\Delta T$ leads to finding a local criterion by solving Eq.(\ref{cond max dt 1}), where the functional derivative of $\Delta T$ by $\alpha (x)$ has to be considered. The local criterion obtained takes the form of a condition on $\frac{\partial \rho  }{\partial \alpha}(x)$ where $\rho (\alpha )$ is material dependent.

\begin{eqnarray}
\frac{\partial \Delta T }{\partial \alpha (x)} = 0 
\label{cond max dt 1}
\end{eqnarray}

\begin{eqnarray}
A  \frac{\partial B}{\partial \alpha (x)} = B  \frac{\partial A}{\partial \alpha (x)} 
\label{cond max dt 2}
\end{eqnarray}

\begin{eqnarray}
\frac{\partial B}{\partial \alpha (x)} =  \int_0^L  \int_0^v   \frac{\partial \rho (u)}{\partial \alpha (x)}  \, \mathrm du \, \mathrm dv       
\end{eqnarray}

\begin{eqnarray}
\frac{\partial A}{\partial \alpha (x)} =  2 \left( \int_0^L \frac{\partial \alpha (v)}{\partial \alpha (x)}  \mathrm dv  \right)  \left( \int_0^L \alpha(v) \, \mathrm dv  \right)    
\end{eqnarray}

So if $x > v$, $x$ is not in the $[0,v]$ interval :

\begin{eqnarray}
\int_0^v \frac{\partial \rho (u)}{\partial \alpha (x)}  \, \mathrm du = 0
\label{cond max dt 3_3}
\end{eqnarray}

And if $x < v$, $x$ is in the $[0,v]$ interval :

\begin{eqnarray}
\int_0^v \frac{\partial \rho (u)}{\partial \alpha (x)}  \, \mathrm du = \frac{\partial \rho (x)}{\partial \alpha (x)}  \int_0^L \delta_{x} (u) \, \mathrm du
\label{cond max dt 3_4}
\end{eqnarray}

\begin{eqnarray}
\frac{\partial B}{\partial \alpha (x)} = \left(L-x\right) \frac{\partial \rho}{\partial \alpha}(x) \int_0^L \delta_{x} (u) \, \mathrm du  
\label{cond max dt 3_5}
\end{eqnarray}

\begin{eqnarray}
\frac{\partial A}{\partial \alpha (x)} =  2 \left( \int_0^L \delta_{x} (u) \, \mathrm du   \right)  \left( \int_0^L \alpha(v) \, \mathrm dv  \right)        
\label{cond max dt 3_6}
\end{eqnarray}

With $ \delta_{x} $ the Dirac delta function centered on $x$. Using Eqs. (\ref{cond max dt 2},\ref{cond max dt 3_5}) and (\ref{cond max dt 3_6}), we get :

\begin{eqnarray}
\frac{\partial \rho  }{\partial \alpha}(x) = \frac{1}{1-\frac{x}{L}} \frac{\partial \rho  }{\partial \alpha}(0)
\label{cond max dt 4}
\end{eqnarray}

Eq. (\ref{cond max dt 4}) gives a local criterion of an optimized graded thermoelectric cooler.

\section{Discussion about the equation of state}

From Eq. (\ref{cond max dt 4}) we deduce that the optimization strongly relies on the relation between the Seebeck coefficient and the electronic conductivity. This relation depends on the equation of state of the electron gas that is considered. As an example, we used a non-degenerated Lorentz Gas equation of state to evaluate the maximum cooling in a semi-conductor, however other equations of state (Price relation for a semiconductor \cite{price1956theory}, exciton \cite{wu2015how}, oxides \cite{walia2013transition}, nanomaterial \cite{humphrey2005reversible}, polymers \cite{li1993granular}) might be used, yielding different improvements of the maximum cooling.

In \cite{bian2006beating} the considered materials have a $zT$ independent of the carrier concentration (and of the electrical conductivity). For any material the $zT$ parameter depends on the carrier concentration, this will impact the maximum cooling temperature.

\begin{figure}
\includegraphics[width=0.48\textwidth]{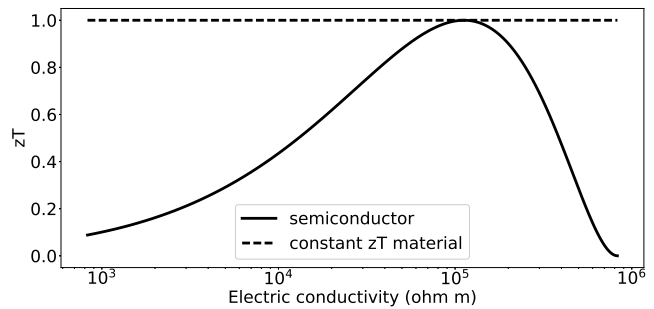}
\caption{Figure of merit as a function of the electric conductivity for a semiconductor (straight line) and for a hypothetical constant $zT$ material (dash line). The maximum $zT$ of the semiconductor is equal to the $zT$ of the constant $zT$ material, which means that for a homogeneous material the maximum temperature difference will be the same.}
\label{fig zt}
\end{figure}

\begin{figure}
\centering
\includegraphics[width=0.48\textwidth]{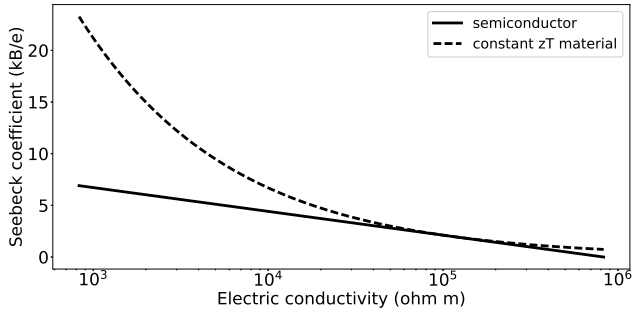}
\caption{Seebeck coefficient as a function of the electric conductivity for a semiconductor (straight line) and for a constant $zT$ material model (dash line). At any electrical conductivity the Seebeck coefficient is higher in the case of a constant $zT$ material model with equality with the semiconductor model for a value Seebeck of $2k_{B}/e$, which corresponds to the value where the maximum $zT$ is reached for the semiconductor model. For low carrier density (low electric conductivity) in a real semiconductor the Seebeck coefficient reaches a maximum, this effect is not present in our simple model.}
\label{fig alpha}
\end{figure}

A numerical investigation with experimental properties of $\mathrm B \mathrm i_{2} \mathrm T \mathrm e_{3}$ yields a $27\%$ increase \cite{bian2007maximum} and a $35\%$ increase is predicted with silicon \cite{bian2006cooling}. For comparison between the semiconductor model we investigated and the constant $zT$ material that Bian et al. used, we plot  the figure of merit (in Figure \ref{fig zt}) and the Seebeck coefficient (in Figure \ref{fig alpha}) as functions of the electric conductivity. One consequence drawn from this work is that an upper limit is established for the possible cooling temperature.

\section{Application for a semi-conductor}

For the sake of simplicity we apply the previous development to a non-degenerated Lorentz gas. It is possible to apply this relation to more complex systems, however the model we choose is a simple model for the thermoelectric properties of a semi-conductor. The electrical conductivity and the Seebeck coefficient are functions of the doping $n$.

\begin{eqnarray}
\sigma =  n  e \mu 
\label{rho func n}
\end{eqnarray}

$\mu$ denotes the electron mobility and $e$ is the elementary charge.

\begin{eqnarray}
\alpha = - \frac{k_{B}}{e} \ln \left( \frac{n}{N_{eff}}  \right) 
\label{seebeck func n}
\end{eqnarray}

Based on Eq. (\ref{rho func n}) and (\ref{seebeck func n}) we obtain a relation between the electrical resistivity and the Seebeck coefficient with $N_{eff}$ the effective density of states (as defined in \cite{ioffe1957semiconductor}). This model is valid for a non degenerated semiconductor.

\begin{eqnarray}
\rho = \rho _{0} \exp \left(  \frac{e  \alpha}{k_{B}}  \right) 
\label{rho func seebeck}
\end{eqnarray}

From Eq. (\ref{rho func seebeck}) and  the local criterion (Eq. (\ref{cond max dt 4})) we obtain the resistivity and the Seebeck coefficient as functions of the position $x$.

\begin{eqnarray}
\rho (x) = \rho (0) \frac{1}{1-\frac{x}{L}}
\label{rho func x}
\end{eqnarray}

\begin{eqnarray}
\alpha (x) = \alpha (0) + \frac{k_{B}}{e} \ln \left( \frac{1}{1-\frac{x}{L}} \right) 
\label{seebeck func x}
\end{eqnarray}

The obtained solution diverges at the hot side for both the electrical resistivity and the Seebeck coefficient. The electrical conductivity is a linear function of the position and takes the value zero at the hot side. This solution gives a hypothetical solution. However for more realistic situations we investigate solutions where the electrical conductivity is a linear function of the position and is not equal to zero on the hot side. 

Using Eq. (\ref{rho func x}) and (\ref{seebeck func x}), we obtain Eq. (\ref{dt func a0 et rho0}), a solution for $\Delta T$ that depends on the properties of the material at the position 0.

\begin{eqnarray}
\Delta T = \frac{T_{0}^{2}}{4 \kappa \rho_{0}}   \frac{    \left(    \alpha (0) +  \frac{k_{B}}{e} \right) ^{2}    }{     \exp  \left( \frac{  e. \alpha (0)}{k_{B}} \right)        }
\label{dt func a0 et rho0}
\end{eqnarray}

The maximization of Eq. (\ref{dt func a0 et rho0}) gives:

\begin{eqnarray}
\alpha (0) = \frac{k_{B}}{e}
\label{a0 pour gradient}
\end{eqnarray}

Which is the traditional prefactor value of any Seebeck expression.

From Eq. (\ref{dt func a0 et rho0}) and (\ref{a0 pour gradient}) the maximal temperature difference in a graded thermoelectric semi-conductor can be computed as:

\begin{eqnarray}
\Delta T _{graded} =  \frac{T_{0}^{2}}{\kappa \rho_{0}}  \frac{    \left(   \frac{k_{B}}{e}    \right) ^{2}       }{   \exp \left( 1 \right)             }
\label{dt max pour gradient}
\end{eqnarray}

A comparison can be made between the graded system and a homogeneous system, the homogeneous case gives a temperature difference of:

\begin{eqnarray}
\Delta T _{homogeneous} =  \frac{T_{0}^{2}}{ 4 \kappa \rho_{0}}  \frac{   \alpha ^{2}       }{   \exp  \left( \frac{  e \alpha }{k_{B}} \right)             }
\label{dt max pour homogene}
\end{eqnarray}

The Seebeck coefficient that gives the maximum of temperature in Eq. (\ref{dt max pour homogene}) is:

\begin{eqnarray}
\alpha_{homogeneous} = 2  \frac{k_{B}}{e}
\label{a0 pour classique}
\end{eqnarray}

The graded system gives a theoretical $36\%$ increase with respect to a classical system. This analytic calculation is coherent with numerical calculations based on experimental material properties\cite{bian2006cooling, bian2007maximum} that yield a theoretical rise of $27\%$ for $\mathrm B \mathrm i_{2} \mathrm T \mathrm e_{3}$ and $35\%$ for $\mathrm S \mathrm i$. The analytic solution diverges at the position $L$ which corresponds to the hot side of the system. Based on Eq. (\ref{eq chaleur fourrier simplifie}), (\ref{rho func x}) and (\ref{seebeck func x}) we compute the temperature as a function of the position in the optimal graded case.

\begin{eqnarray}
\Delta T (x) = \frac{   T_{0}^{2}    \left(  \frac{k_{B}}{e}   \right)^{2}   x   }{    \kappa \rho_{0}  \exp (1)     L     }
\label{dt opt func x}
\end{eqnarray}

Eq. (\ref{dt opt func x}) shows that the temperature is a linear function of the position. At any position the Peltier effect due to the graded material compensates exactly the Joule effect. In this situation the effective heat generation in the graded material is zero and all the heat sources are localized at the interfaces of the graded material.

From Eq. (\ref{rho func x}) we deduce that the electrical conductivity is a linear function of the position which corresponds (based on Eq. (\ref{rho func n})) to a graded material where the doping level is a linear function of the position. For further analysis we studied the maximum cooling of a graded material with a linear doping level. This material will have a doping level of $n_{h}$ at the hot side, $n_{c}$ at the cold side and the doping level is the linear function of the position given by Eq. (\ref{n func x}).

\begin{eqnarray}
n(x) = n_{c} + \left(   n_{h}-n_{c}  \right) \frac{x}{L} 
\label{n func x}
\end{eqnarray}

The ratio $c$ can be defined as $ c = n_{h}/n_{c}$, and this coefficient describes the amplitude of the doping gradient. The value $c=1$ corresponds to a classical system with zero gradient. The electrical resistivity and the Seebeck coefficient can be written as functions of these values at the cold side ($x = 0$) and $c$.

\begin{eqnarray}
\rho (x) = \rho (0) \frac{1}{1-\frac{x}{L}(1-c)}
\label{rho func x c}
\end{eqnarray}

\begin{eqnarray}
\alpha (x) = \alpha (0) + \frac{k_{B}}{e} \ln \left( \frac{1}{1-\frac{x}{L}(1-c)} \right) 
\label{a func x c}
\end{eqnarray}

The Seebeck coefficient at the hot side is:

\begin{eqnarray}
\alpha (L) = \alpha (0) - \frac{k_{B}}{e} \ln \left( c\right) 
\label{a hot}
\end{eqnarray}

\begin{figure}
\centering
\includegraphics[width=0.48\textwidth]{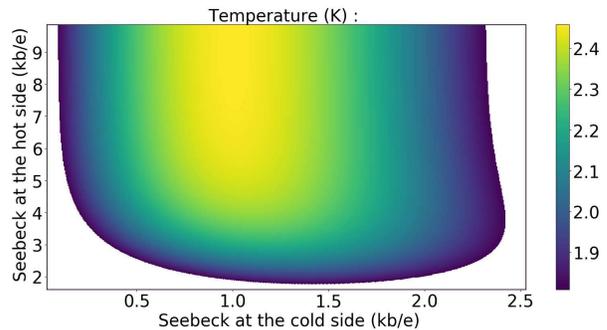}
\caption{Plot of the temperature difference as a function of the Seebeck coefficient at the cold side and the Seebeck coefficient at the hot side. $ \kappa = 10 \mathrm{W.K^{-1} . m^{-1}}$ and $\rho = 10^{-5} \mathrm{\Omega . m}$ This correspond to a semiconductor with a maximum $zT$ of $0.012$. Only the cases where the temperature obtained is superior to the homogeneous optimal case are displayed. In this figure we can notice that there is no particular need of a very high Seebeck coefficient at the hot side to obtain a fair improvement of the temperature difference.}
\label{fid graphe 2d}
\end{figure}

From Eq. (\ref{rho func x c}), (\ref{a func x c}), (\ref{a hot}), (\ref{rho func x}) and (\ref{dt pc max}) we can derive the temperature difference as a function of the Seebeck coefficient at the cold side and the Seebeck coefficient at the hot side. In Figure \ref{fid graphe 2d} we can see that the maximum cooling is obtained when $\alpha_{c} = k_{B}/e$ and $\alpha_{h}$ is as large as possible.

\begin{figure}
\centering
\includegraphics[width=0.48\textwidth]{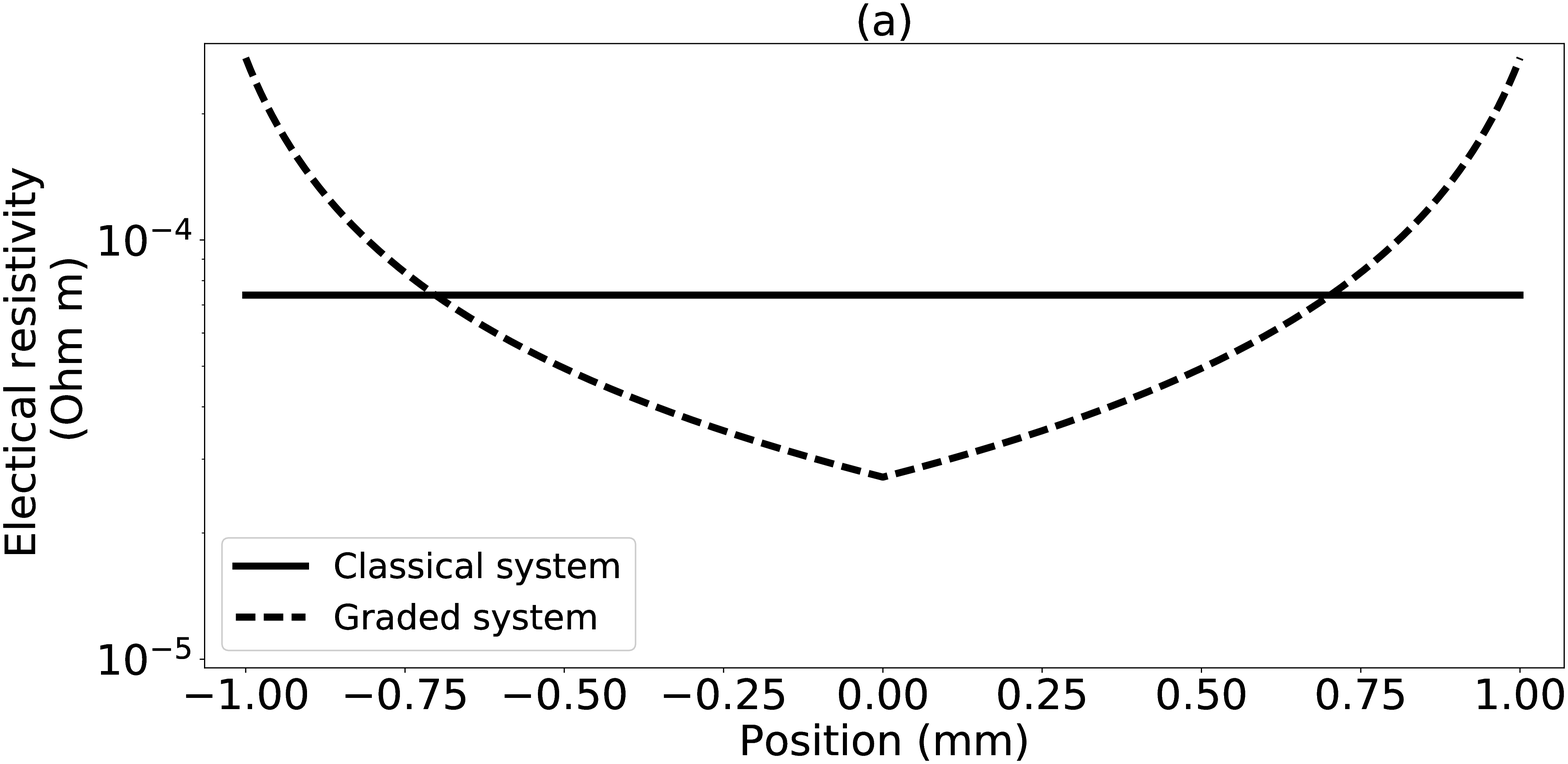}
\includegraphics[width=0.48\textwidth]{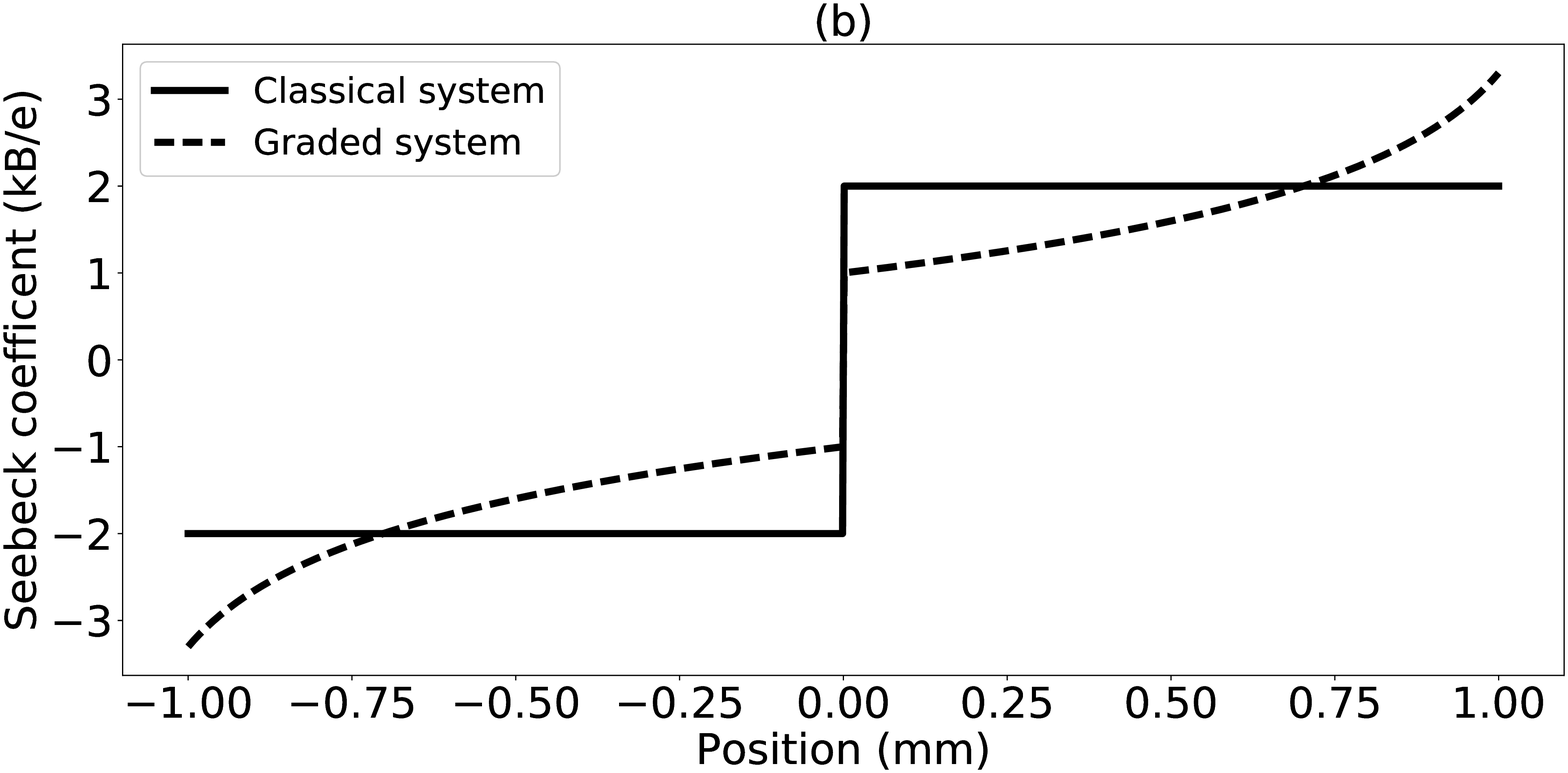}
\includegraphics[width=0.48\textwidth]{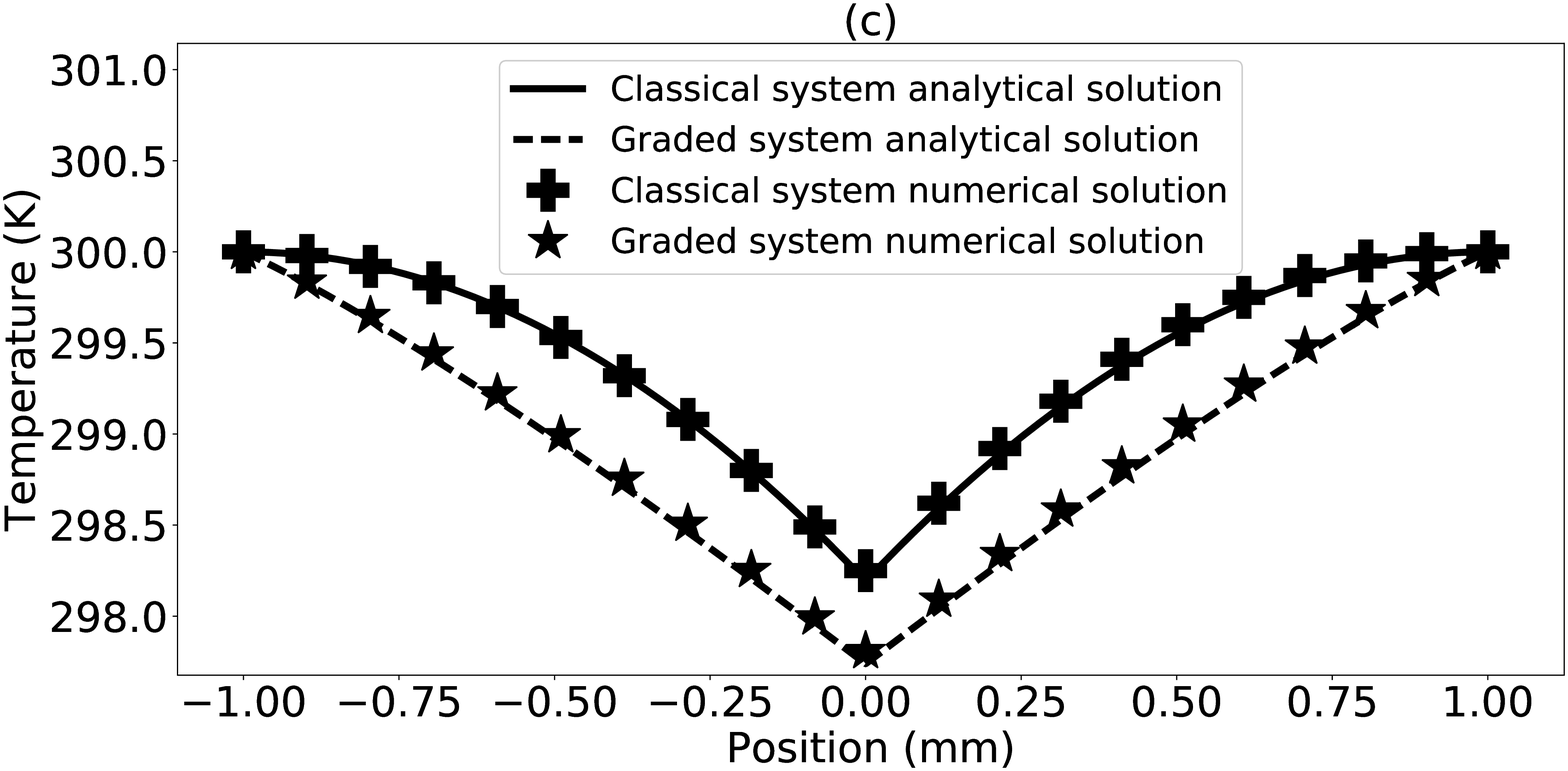}
\caption{For a Seebeck coefficient at the cold side of $k_{B}/e$ and $c = 0.1$ we plot the electrical resistivity (a), the Seebeck coefficient (b) and the temperature (c) as functions of the position, for classical system (straight line) and for graded system (dash line). The analytical and numerical solutions are compared. Numerical solution for classical system is plot with the cross and with stars for the graded solution.}
\label{fig prop func position}
\end{figure}

For a real semiconductor the optimal profile will deviate from our solution at the cold side (due to the decrease in mobility at high carrier density) and at the hot side due to the saturation of the Seebeck coefficient (from bipolar and intrinsic conduction).
In Fig.\ref{fig prop func position} the graded system is obtained for $c=0.1$ which corresponds to a variation of carrier concentration between the hot and cold sides of a factor $10$. For this variation of carrier concentration a clear and fair improvement of the difference of temperature is obtained, these results highlight that there is no need of a important gradient and therefore that deviation from the model at low and high carrier concentration will induce only small consequences on the optimal profile.

\section{Numerical solution for graded thermoelectric system}

A numerical resolution is used to confirm the analytic results without the approximations done for the analytic solution. The numerical solution of a graded material was obtain with the DYCO solver, a numerical solver for coupled equations. This solver is based on a nodal approach and we used it to solve the Onsager relations in a thermoelectric material. In this solver a Millman approach and the conservation of the flux are used. A description of the theoretical background of the DYCO solver can be found in chapter 3 of \cite{goupil2016continuum}. Classical systems and graded systems were solved analytically and numerically under the same conditions.

For the sake of comparison between the analytic solution and the numerical solution, we choose $\kappa=1$, $10$ and $100 \mathrm{W.K^{-1}.m^{-1}}$, $\rho _{0}=10^{-5} \mathrm{Ohm.m}$ and $c =0.1$.

\begin{table*}
\begin{ruledtabular}
\begin{tabular}{ccccc}
Temperature (K) & \multicolumn{2}{c}{homogeneous system} & \multicolumn{2}{c}{inhomogeneous system}\\
\hline 
Solution: &  Analytic & Numeric &  Analytic & Numeric\\ 
\hline 
$\kappa =1 \mathrm{W.K^{-1}.m^{-1}}$& $18.1$ & $15.58$  &   $22.6$ & $19.6$ \\  
$\kappa =10 \mathrm{W.K^{-1}.m^{-1}}$& $1.81$ & $1.75$ &   $2.26$ & $2.19$\\ 
$\kappa =100 \mathrm{W.K^{-1}.m^{-1}}$& $0.181$ & $0.177$  &   $0.226$ & $0.22$\\ 
\end{tabular}
\end{ruledtabular} 
\caption{In this table we summarize results obtained with the resolution of a graded and classical materials with both numeric and analytic resolution. $\kappa = 1, 10, 100$ correspond to $zT = 0.12, 0.012, 0.0012$ respectively.} 
\label{tableau comparaison}
\end{table*}

The analytic solution deviates from the numerical solution for high $zT$ materials which are able to impose a large temperature difference. This is coherent with the approximation made to obtain an analytic solution (we suppose that the temperature stays close to the temperature of the hot side). Numerical solutions confirm that a graded system improves the maximum temperature that can be obtained.

From the solution of the temperature the entropy flux can be evaluate \cite{goupil2011thermodynamics}. This approach can be used to evaluate the different sources of entropy in our system. In order to properly evaluate the entropy sources a solution obtained with no hypothesis is needed. We use the computed numerical solution to perform this evaluation. 

From \cite{goupil2011thermodynamics} the entropy flux is given by :

\begin{eqnarray}
J_{s} = \alpha J - \kappa    \left(     \frac{\frac{\partial T  }{\partial x}}{T} \right) 
\label{entropy flux}
\end{eqnarray}

As we can separate the transport of heat from convection and conduction \cite{apertet2012internal}, the entropy flux can be separated in convection (first term of Eq. (\ref{entropy flux})) and conduction (second term of Eq. (\ref{entropy flux})).

The variation of entropy flux is :

\begin{eqnarray}
\nu_{f} = \frac{\partial J_{s}  }{\partial x} =  J \frac{\partial \alpha }{\partial x}- \kappa  \frac{\partial  \left(     \frac{\frac{\partial T  }{\partial x}}{T} \right) }{\partial x} 
\label{variation entropy flux}
\end{eqnarray}

In stationary condition the produced entropy ($\nu_{c}$) is equal to the variation of entropy flux ($\nu_{f}$). If we consider the heat equation (Eq. (\ref{eq chaleur fourrier})), the entropy produced is given in Eq. (\ref{entropy created}). The entropy produced is composed of two terms, a term related to the Joule heating and a term related to the thermal gradient. 

\begin{eqnarray}
\nu_{c} = \nu_{f}
\end{eqnarray}

\begin{eqnarray}
\nu_{c} =  J \frac{\partial \alpha }{\partial x} -  \kappa \frac{\partial^{2} T }{\partial x^{2}} + \kappa \frac{ \left(  \frac{\partial T }{\partial x} \right)^{2} }{T^{2}}
\end{eqnarray}

\begin{eqnarray}
\nu_{c} =  \frac{ \rho J^{2} }{T}  + \kappa \frac{ \left(  \frac{\partial T }{\partial x} \right)^{2} }{T^{2}}
\label{entropy created}
\end{eqnarray}

\begin{figure}
\centering
\includegraphics[width=0.48\textwidth]{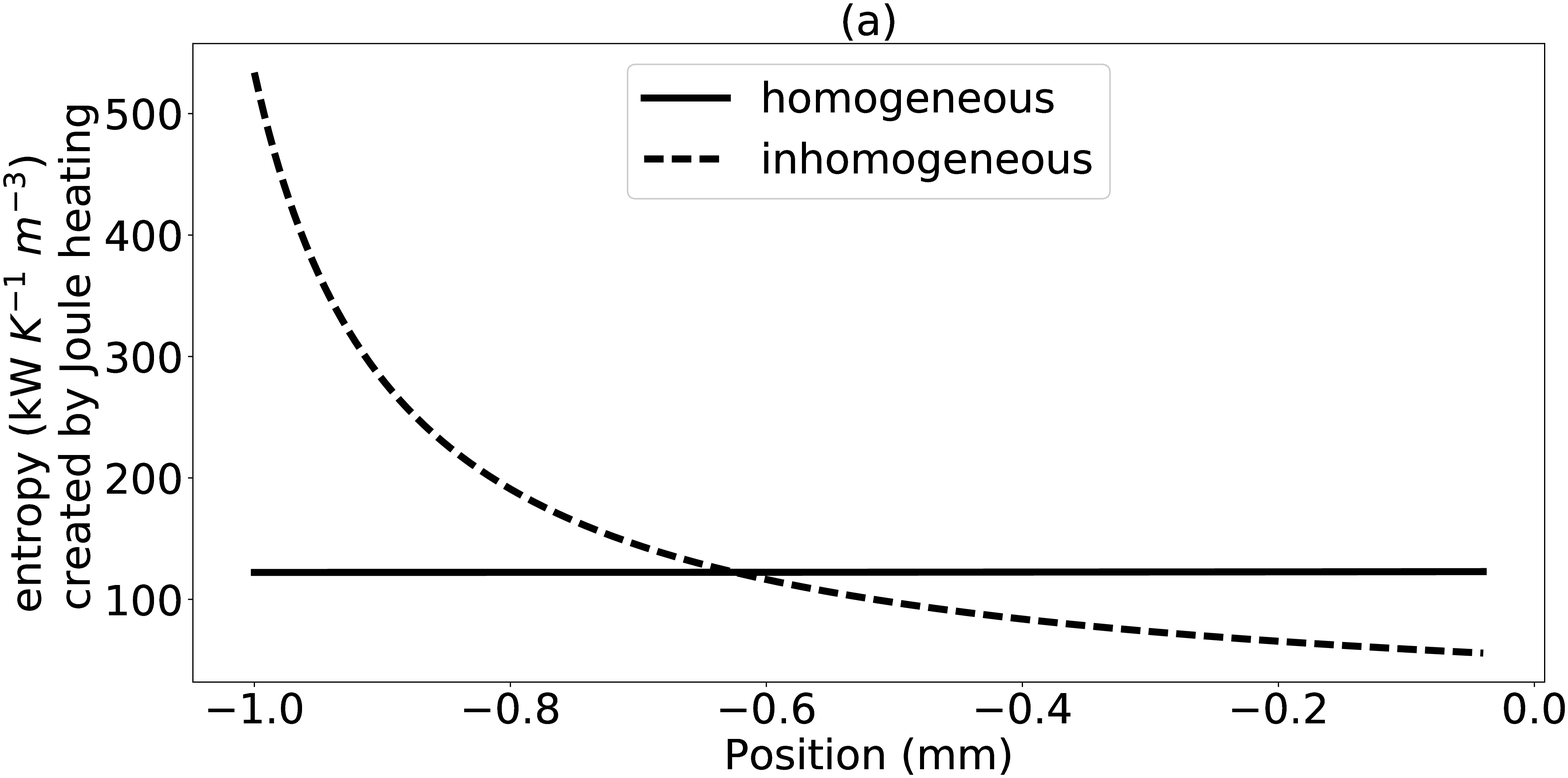}
\includegraphics[width=0.48\textwidth]{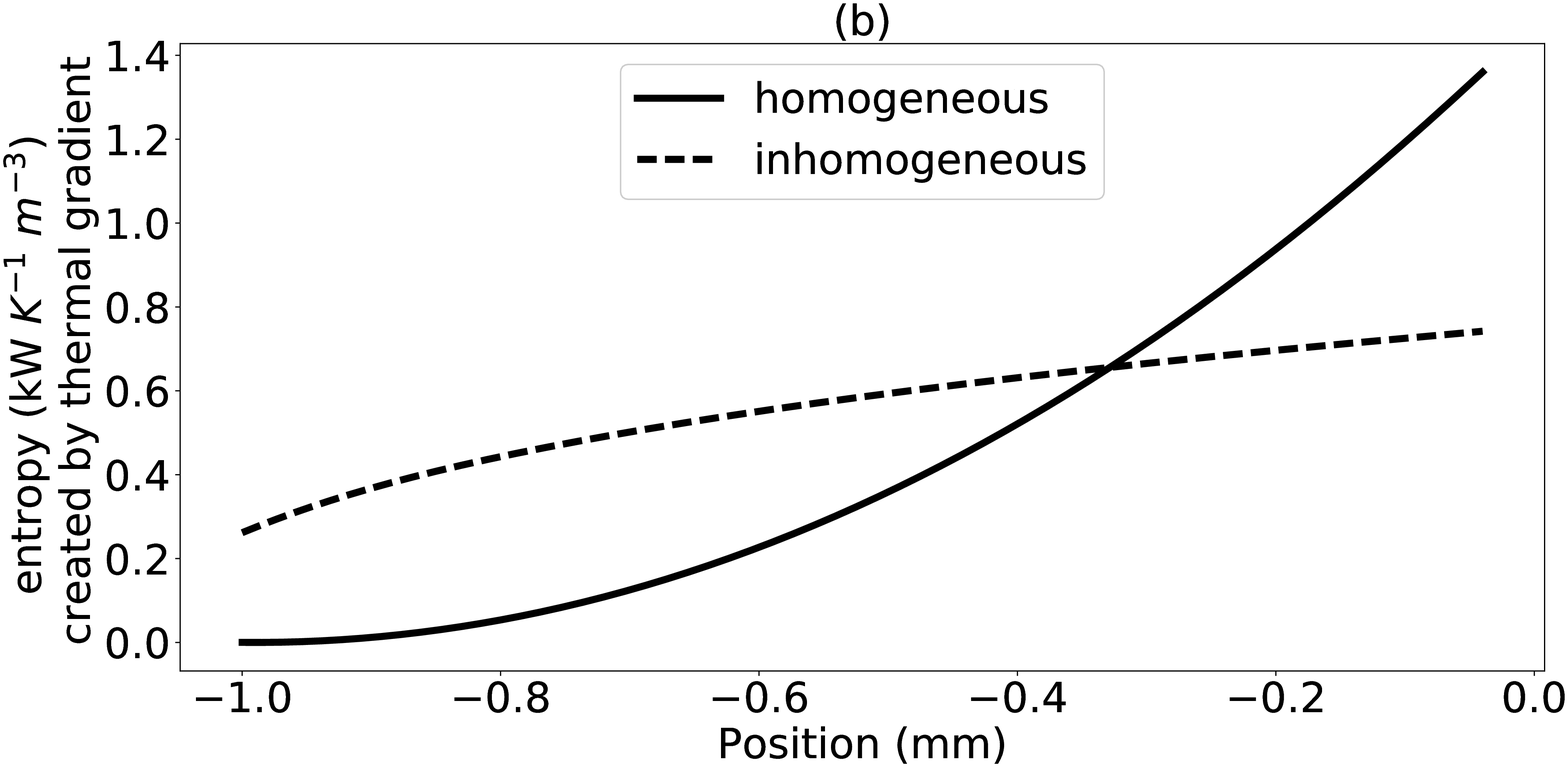}
\caption{Plot of the entropy produced between the hot side ($x=-1$mm) and the cold side ($x=0$). The entropy produced by Joule heating (a) is higher than the entropy produced by the thermal gradient (b). For both entropy sources the entropy produced is higher at the hot side for the inhomogeneous (dash line) case and higher at the cold side for the homogeneous case (straight line).}
\label{fig entropy cree}
\end{figure}

In figure \ref{fig entropy cree} we show that the produced entropy is lower at the cold side in the inhomogeneous case. This lower entropy production is mainly due to the lower Joule heating at the cold side in the inhomogeneous case. 

The average entropy produced by Joule effect is $122 \mathrm{kW.K^{-1}.m^{-3}}$ for the homogeneous case and $140 \mathrm{kW.K^{-1}.m^{-3}}$ for the inhomogeneous case. The average entropy produced by the thermal gradient effect is $0.45 \mathrm{kW.K^{-1}.m^{-3}}$ for the homogeneous case and $0.56 \mathrm{kW.K^{-1}.m^{-3}}$ for the inhomogeneous case. For both sources of entropy the total entropy produced is higher in the inhomogeneous case.

The total entropy produced by the system is rejected in the thermostat at both hot sides. This can be obtained by integrating $\nu_{c}$ over the entire device. 

\begin{eqnarray}
V_{c} = \int_{-L}^L \left(  \frac{ \rho J^{2} }{T}  + \kappa \frac{ \left(  \frac{\partial T }{\partial x} \right)^{2} }{T^{2}} \right)  \, \mathrm dx 
\end{eqnarray}

\begin{eqnarray}
V_{c} = J (\alpha (L) - \alpha (-L))
\nonumber\\
- \kappa \left(     \frac{\frac{\partial T  }{\partial x}(L)}{T(L)} - \frac{\frac{\partial T  }{\partial x}(-L)}{T(-L)}\right)
\label{total entropy created}
\end{eqnarray}

From figure \ref{fig entropy cree} we observe that the produced entropy is mainly due to the Joule heating. In the graded system the produced entropy is higher at the hot side due to higher Joule heating.  

The graded case allows higher performance for the thermoelectric cooler at the cost of a higher entropy production. The improved performance are obtained through a redistribution of the entropy production, the graded system has a lower entropy production at the cold side at a cost of a higher total entropy production.

\section*{Conclusion}

We analyzed a FGM-based Peltier cooler by both analytic and numerical means. Both yield an improvement of the temperature difference by using graded materials. The analytic solution of the heat equation shows that a local criterion can be found in order to maximize the temperature difference. This criterion was used within an analytic model of a thermoelectric semi-conductor. This shows that it should be possible to improve by $36\%$ the maximum of temperature with an optimized graded semiconductor, which corresponds to a cooling down of $-88 \mathrm{K}$, as compared with only  $-65 \mathrm{K}$ reached with a homogeneous material \cite{huang2000design}. This improvement is equivalent to other theoretical evaluations based on real material properties \cite{bian2006cooling, bian2007maximum}.

The improvement of the temperature difference through graded materials has been confirmed with numerical calculations and shows that the hypothesis used for the analytical analysis yields some overestimation of the cooling effect. The overestimation of the cooling effect is due to the approximation made ($J T(x) \frac{\partial \alpha}{\partial x}(x) = JT_{0}\frac{\partial \alpha}{\partial x}(x)$) that leads to an overestimation of the Thomson-Peltier effect since the temperature is lower than $T_{0}$.

The redistribution of Joule and Peltier effects increases the maximum cooling. In the ideal case the sum of the Peltier cooling due to the graded material and the Joule heating is zero. In this situation the temperature displays a linear profile and the maximum cooling is reached. This improvement highly depends on the variation of the material properties with the doping, so other types of material might give higher possible improvements. An entropy creation analysis through numerical computation shows that the Joule effect is the main source of entropy and that the entropy creation is lower on the cold side in the optimized graded case. The graded system forces the entropy production to be localized on the hot side which increases the maximum cooling at a cost of a higher total entropy production (leading to a lower efficiency).  

\begin{acknowledgments}
The authors thank ANRT (CIFRE) for the funding of doctoral studies by E. Thiebaut and ST Microelectronics TOURS for their support.
\end{acknowledgments}

\end{document}